\documentclass{aastex}
\usepackage{amsmath}
\usepackage{graphicx}
\usepackage{url}
\usepackage{amssymb}
\usepackage{float}
\makeatletter
\let\@dates\relax
\makeatother

\shortauthors{A. BRUNO}

\shorttitle{CALIBRATION OF GOES PROTON DETECTORS WITH PAMELA SEP OBSERVATIONS}

\begin{document}

%
%

\title{Calibration of the GOES-13/15 high energy proton detectors based on the PAMELA solar energetic particle observations}

%
%

\author{A. Bruno}
\affil{Istituto Nazionale di Fisica Nucleare (INFN), Sezione di Bari, I-70126 Bari, Italy.}
\affil{alessandro.bruno@ba.infn.it.}
\affil{\it Accepted for publication in Space Weather}

%
%

\begin{abstract}
The EPEAD and HEPAD instruments on the GOES spacecraft have served over many years as monitors of the solar particles intensities, surveying the Sun and measuring in situ its effect on the near-Earth solar-terrestrial environment. However, the reconstruction of the differential energy spectra is affected by large uncertainties related to the poor energy resolution, the small geometrical factor and the high contamination by 
out-of-acceptance particles.
In this work, the high quality data set from the PAMELA space mission is used to calibrate the high energy ($>$80 MeV) proton channels of the EPEAD and the HEPAD sensors onboard the GOES-13 and -15, bringing the measured spectral intensities in-line with those registered by PAMELA. Suggested corrections significantly reduce the uncertainties on the response of GOES detectors, thus improving the reliability of the spectroscopic observations of solar energetic particle events.
\end{abstract}

%
%

\section{Introduction}
The multi-mission Geostationary Operational Environmental Satellite (GOES) program is a joint effort of NOAA and NASA aimed to the monitoring of the near-Earth space, including operational meteorology and space weather.
GOES weather imagery and quantitative sounding data offer a continuous and reliable stream of environmental information used to support weather forecasting, severe storm tracking and meteorological research.
In particular, the GOES data are vital to the space weather monitoring, surveying the Sun and measuring in situ its effects on the near-Earth environment.
The GOES Space Environment Monitor (SEM) system contains several subsystems including three instruments
designed for particle flux observations:
two Energetic Proton, Electron, and Alpha Detectors (EPEADs), pointing to East and to West respectively; one High Energy Proton and Alpha Detector (HEPAD), pointing to the zenith.
The EPEADs measure protons in the energy range 0.74--900 MeV, alpha particles in the energy range 3.8--500 MeV, and
electrons in three energy ranges $>$0.6, $>$2 and $>$4 MeV. The HEPAD registers protons and alpha particles with energies $>$330 MeV and $>$2.56 GeV, respectively.

The measurement of the energetic spectra provides fundamental information to constraint the models of Solar Energetic Particle (SEP)
origin and propagation
in the interplanetary space (see e.g. \citet{ref:REAMES}). Consequently, accounting for the relatively poor resolution,
the knowledge of the uncertainties affecting the estimate of SEP
differential fluxes based on the nominal energetic response of the different channels is a crucial ingredient in the calculation \citep{ref:SmartShea1999}.
Recently, \citet{ref:Rodriguez2014} carried out a relative intercalibration work using the proton channels of the Energetic Particles Sensors (EPS) on board different GOES spacecraft, based on
solar wind dynamic pressure criteria.
In addition, \citet{ref:Sandberg2014} (hereafter referred as S14) presented a calibration study of the EPS channels from units
onboard GOES-5, -7, -8, and -11 using as reference the science-level NASA IMP-8/Goddard Medium Energy (GME) experiment data set.
The cross-calibrated proton energies
derived for the P2--P5 channels ($\lesssim$80 MeV) were lately validated by \citet{ref:Rodriguez2017} by comparison with the Solar Terrestrial Relations Observatory (STEREO) data.
In this work, the high energy ($\gtrsim$80 MeV) SEP measurements of the Payload for Antimatter Matter Exploration and Light-nuclei Astrophysics (PA\-ME\-LA) are used to calibrate the proton channels of the EPEAD (P6--P7) and the HEPAD (P8--P11) sensors onboard GOES-13 and -15.

\section{PAMELA data}\label{PAMELA data}
PAMELA is a space-based experiment designed for a precise measurement of charged Cosmic-Rays (CRs) -- protons, electrons, their antiparticles and light nuclei -- in the energy interval from several tens of MeV up to several hundreds of GeV \citep{ref:PHYSICSREPORTS}.
The instrument consists of a magnetic spectrometer equipped with a si\-li\-con tracking system, a time-of-flight system shielded by an anticoincidence system, an electromagnetic calorimeter and a neutron detector.
The Resurs-DK1 satellite, which hosts the apparatus, was launched into a semi-polar (70 deg inclination) and elliptical (350--610 km altitude) orbit on 15 June 2006; in 2010 it was changed to an approximately circular orbit at an altitude of $\sim$ 580 km.

PA\-ME\-LA is providing comprehensive observations of the interplanetary and magnetospheric radiation in the near-Earth environment (see, e.g., \citet{ref:SOLARMOD,ref:TRAPPED,ref:ALBEDO,ref:GSTORM,ref:BRUNO_ICRC2015a,ref:BRUNO_ICRC2015b}).
In particular, PA\-ME\-LA is able to precisely measure the SEP events during the solar cycles 23 and 24, in a wide energy interval encompassing the low energy observations by in-situ instruments and the Ground Level Enhancement (GLE) data by the worldwide network of neutron monitors \citep{ref:SEP2006,ref:MAY17PAPER,ref:BRUNO_ICRC2015c,ref:BRUNO_JPCS}.
Full details about apparatus performance, proton selection, detector efficiencies and experimental uncertainties can be found in \citet{ref:SOLARMOD,ref:PHYSICSREPORTS}.

The list of the SEP data used in this work, including 24 major events detected by PA\-ME\-LA prior to 2014 October, is reported in Table \ref{tab:used_events}. Since the present calibration work is focused on GOES-13 and -15 detectors (whose data are available starting from 2010 May and 2011 January, respectively),
the 2006 December 13 and 14 SEP events \citep{ref:SEP2006} were excluded from the selected data sample -- a calibration study of the GOES-10/11/12 proton detectors, based on the aforementioned events, will be the object of a forthcoming publication.
So far, a relatively low rate of energetic events has been registered during the current solar cycle, with the most relevant including the 2012 May 17 GLE
\citep{ref:Gopalswamy2013,ref:MAY17PAPER,ref:BRUNO_JPCS} and the 2014 January 6 (sub?)GLE \citep{ref:Thakur2014}.
Finally, it should be noted that the used SEP list
includes
four pairs of overlapping events:
the 2011 September, the 2012 January, the 2012 July and the 2014 January ones; consequently, the spectra measured for the second event of each pair comprises a contribution from the first event.

\begin{table}[h]
    \centering\small
    \begin{tabular}{c|c|c|c|c|c|c}
     \multicolumn{2}{c}{} & \multicolumn{2}{|c}{GOES-13} & \multicolumn{2}{|c}{GOES-15} & \multicolumn{1}{|c}{} \\
    \hline
    \# & Event & A & B & A & B & P$_{max}$\\
    \hline
    1 & 2011 Mar 21 & E & W & W & E & P10\\
    2 & 2011 Jun 07 & E & W & W & E & P10\\
    3 & 2011 Sep 06 & E & W & E & W & P9\\
    4 & 2011 Sep 07 & E & W & E & W & P9\\
    5 & 2011 Nov 04 & E & W & E & W & P8\\
    6 & 2012 Jan 23 & E & W & E & W & P7\\
    7 & 2012 Jan 27 & E & W & E & W & P10\\
    8 & 2012 Mar 13 & E & W & W & E & P10\\
    9 & 2012 May 17 & E & W & W & E & P11\\
    10 & 2012 Jul 07 & E & W & W & E & P8\\
    11 & 2012 Jul 08 & E & W & W & E & P9\\
    12 & 2012 Jul 19 & E & W & W & E & P7\\
    13 & 2012 Jul 23 & E & W & W & E & P7\\
    14 & 2013 Apr 11 & E & W & W & E & P8\\
    15 & 2013 May 22 & E & W & W & E & P9\\
    16 & 2013 Sep 30 & E & W & E & W & P7\\
    17 & 2013 Oct 28 & E & W & E & W & P7\\
    18 & 2013 Nov 02 & E & W & E & W & P7\\
    19 & 2014 Jan 06 & E & W & E & W & P11\\
    20 & 2014 Jan 07 & E & W & E & W & P9\\
    21 & 2014 Feb 25 & E & W & E & W & P10\\
    22 & 2014 Apr 18 & E & W & W & E & P7\\
    23 & 2014 Sep 01 & E & W & W & E & P10\\
    24 & 2014 Sep 10 & E & W & W & E & P7\\
    \hline
    \end{tabular}
    \caption{\small{List of the 24 SEP events observed by PAMELA and used in this work. The orientation (East/West) of the two EPEAD sensors (A/B) is reported for both GOES-13 and -15. The last column reports the highest GOES energy channel in which a discernible SEP signal was detected.}}
    \label{tab:used_events}
\end{table}

Due to the shielding effect of the Earth's magnetosphere, low rigidity (R = momentum / charge) interplanetary CRs can be detected only when the satellite passes through relatively high magnetic latitude regions, where the corresponding geomagnetic cutoff rigidity is lower than the PAMELA proton threshold ($\sim$400 MV, constrained by trigger requirements). In order to discard trapped/albedo particles and avoid geomagnetic effects \citep{ref:BRUNO_HAWAII},
fluxes are conservatively estimated by selecting protons with rigidity 1.3 times higher than the local vertical St\"ormer cutoff.
Consequently, the duty cycle relative to the orbital period is smaller for lower energy protons;
it also varies with the geographic longitude due to the asymmetries between the terrestrial rotational and magnetic axes.

The background associated to Galactic CRs (hereafter GCRs) is evaluated by using the average proton flux measured by PAMELA during the 24 hours prior to the SEP event onset -- for sake of consistency with the correction applied to GOES intensities (see Section \ref{Data analysis}) -- and subtracted from the total spectra.
In case of multiple events, the GCR component estimated for the first event is also used for the second one.
Pitch angle anisotropies with respect to the local interplanetary magnetic field direction are accounted for by estimating the instrument ``asymptotic'' exposition along the satellite orbit (see \citet{ref:BRUNO_ICRC2015c,ref:BRUNO_JPCS} for details).

\section{GOES data}\label{Section:GOES data}
Currently, concerning solar particle observations, the GOES system consists of the GOES-13 and -15 units (3$^{rd}$ generation program), launched on 2006 May and 2010 March, and operating as GOES-East (75$^{\circ}$W) and GOES-West (135$^{\circ}$W), respectively.
GOES-13/15 are capable of a yaw flip, in which the spacecraft rotates about the axis pointed toward the Earth's center so that the two EPEAD-A/B sensors invert their orientation. In particular, GOES-15 undergoes a yaw flip twice a year at the equinoxes, while the GOES-13/EPEAD orientation is
fixed since it went operational \citep{ref:Rodriguez2014}.
The EPEAD-A/B looking directions during the SEP events used in this work are reported in Table \ref{tab:used_events}. The differences between the flux intensities measured by east/westward viewing detectors can be related to East-West effects \citep{ref:Rodriguez2010}. However, such effects are not relevant at PA\-ME\-LA energies
because of the relatively large gyro-radius of interplanetary protons entering the Earth's magnetosphere.
Consequently, the observations from both EPEAD-A/B can be compared to PA\-ME\-LA measurements.

The EPEADs are simple solid-state sensors based on pulse-height discrimination, including 7 proton energy channels (P1--P7) covering the interval from 0.74 to 900 MeV (nominal range). In particular, the two highest energy channels (P6--P7) overlap the PA\-ME\-LA range (see Table \ref{tab:nominal_ranges}). The detectors were designed to handle large count rates without overwhelming the electronics. However, since they were built with passive shielding (no anti-coincidence system), measurements are affected by significant side and rear penetration effects, i.e. particles can pass through the shielding from any direction and be counted as though they had entered through
the nominal detector entrance aperture.
An algorithm (R. Zwickl, unpublished note, 1989; see Appendix A in \citet{ref:Rodriguez2017} for a comprehensive description) was developed in order to correct particle fluxes in real time (hereafter the EPEAD data based on the Zwickl algorithm are denoted as Z89 to avoid ambiguities). The algorithm accounts for the response of the EPEAD channels in terms of a primary energy range and multiple secondary (i.e. spurious or contamination) energy ranges.
It should be noted that the implemented
correction assumes a power-law spectrum with a $\gamma$ = 3 spectral index; consequently, the Z89 fluxes can over/underestimate the SEP ``true'' intensities if the spectral shape deviates significantly \citep{ref:Vainio1995}.
Both uncorrected and Z89 fluxes are available at the NOAA archive (\url{http://www.ngdc.noaa.gov/stp/satellite/goes/dataaccess.html}).

The HEPADs consist of telescope assemblies with two solid-state sensors along with a Cerenkov radiator/PMT providing directional (front/rear incidence) discrimination and energy selection \citep{ref:Sellers1996}. They comprise four proton energy channels (P8--P11) whose nominal ranges are reported in Table \ref{tab:nominal_ranges}.
In the case of the integral channel P11 ($>$700 MeV),
data were processed by NOAA as a differential channel with energy integrated geometrical factor $\int G(E) dE$ = 1565 cm$^{2}$ sr MeV and an effective energy of 1000 MeV (see Sauer corrections documented by \citet{ref:SmartShea1999}).
HEPAD major weaknesses include the small geometrical factor ($\sim$0.73 cm$^{2}$sr), the poor energy resolution and the high contamination from out-of-acceptance particles (especially rear-penetration effects).

\begin{table}[!t]
\centering
\begin{tabular}{r||c|c||c|c|c|c}
\multicolumn{1}{r||}{} & \multicolumn{2}{c||}{EPEAD} & \multicolumn{4}{c}{HEPAD} \\
\hline
Channel & P6 & P7 & P8 & P9 & P10 & P11 \\
\hline
Energy range (MeV) & 84 - 200 & 110 - 900 & 330 - 420 & 420 - 510 & 510 - 700 & $>$ 700 \\
Mean energy (MeV) & 165 & 433 & 375 & 465 & 605 & 1000? \\
\hline
\end{tabular}
\caption{\small{Nominal energy ranges and mean energies of the EPEAD (P6, P7) and the HEPAD (P8, P9, P10, P11) channels. Provided mean energies are valid for flat spectra.}}
\label{tab:nominal_ranges}
\end{table}

\section{Data analysis}\label{Data analysis}
The method used to calibrate the GOES proton detectors with PA\-ME\-LA data is based on the algorithm developed by S14 to calibrate the EPS units onboard GOES-11 and previous missions by means of the IMP-8/GME data. It consists of the application of an iterative linear regression scheme between PA\-ME\-LA and GOES SEP fluxes with identical time stamp.
In contrast to S14, the background in GOES data due to contaminations and GCRs is excluded from the calculation.
As aforementioned, the calibration procedure is based on the SEP observations made by PA\-ME\-LA during a relatively large time interval (2011--2014). Consequently, the inclusion of the background, not negligible for the highest energy channels (P6--P11) and with time-dependent variations related to the solar activity, would lead to systematical effects in the assessment of the channel ``effective'' energies. Both uncorrected and Z89 EPEAD intensities are used in the calculation. For the former, the background is evaluated by using uncorrected intensities averaged over the 24 hours prior to the SEP arrival (quiet solar periods); in case of overlapping events, the background measured before the first one is used. An analogous correction is also applied to the HEPAD fluxes.
The residual background in the Z89 fluxes is assumed to be insignificant.

\subsection{EPEAD calibration}\label{EPEAD calibration}
The procedure developed to calibrate the EPEAD sensors is structured as follows.
\begin{itemize}
\item PA\-ME\-LA SEP fluxes are evaluated on a relatively short timescale ($\sim$48 minutes), corresponding to spacecraft semi-orbits; the same time resolution is used for the EPEADs by averaging the 5-minute intensities (\url{https://satdat.ngdc.noaa.gov/sem/goes/data/new_avg/}). Time bins characterized by anisotropic flux distributions (e.g. during the onset phase of the 2012 May 17 GLE event \citep{ref:MAY17PAPER,ref:BRUNO_JPCS}) are excluded.
\item The PA\-ME\-LA energy spectra are first estimated by using 9 logarithmic bins covering the range 80 -- 370 MeV. The mean energies are computed according to \citet{ref:LaffertyWyatt}, assuming a power-law spectrum with a $\gamma$ = 3 spectral index:
    \begin{equation}\label{LaffertyWyatt_Eq}
    E_{mean}=\left[\frac{E_{max}^{1-\gamma}-E_{min}^{1-\gamma}}{(E_{max}-E_{min})\cdot(1-\gamma)}\right]^{-\frac{1}{\gamma}},
    \end{equation}
    where $E_{max}$ and $E_{min}$ are the upper and the lower energy limits of the considered bin. Deviations from the $\gamma$ = 3 spectrum are found to be negligible ($\Delta E_{mean}\lesssim$0.3\% for 1$<\gamma<$6) due to the relatively small bin widths.
\item Then, exploiting the superior energy resolution of the PA\-ME\-LA instrument,
    each bin is subdivided into an ultradense logarithmic grid (400 values) and fluxes are interpolated at the energy $E_{j}$ of each sub-bin $j$ based on a power-law fit of PA\-ME\-LA spectra below 350 MeV, creating 9$\times$400 EPEAD versus PA\-ME\-LA flux distributions (for both P6 and P7 channels), using the data points from all the selected SEP events (see Table \ref{tab:used_events}).
\item For each EPEAD channel ($i$=$6,7$), distributions are fitted with a linear function:
    \begin{equation}\label{eq:linear_fit}
    F_{GOES,i} = a_{i,j} \cdot F_{PAM}\left(E_{j}\right),
    \end{equation}
    where $a_{i,j}$ (free parameter) represents the scaling factor at energy $E_{j}$.
\item The fit accounts for uncertainties on both PA\-ME\-LA and EPEAD fluxes. The one-sigma error band relative to the power-law fit (propagating parameters uncertainties) is associated to the PAMELA points, including a 20\% systematical error (selection efficiencies, background estimate, etc.).
    The statistical uncertainties are evaluated by accounting for the GCR background subtraction, by using 68.27\% confidence level intervals for a Poisson signal $F_{tot}$ in presence of a background $F_{gcr}$ \citep{ref:FeldmanCousins}.
    The number of protons registered by each energy channel is derived by exploiting the count rate information available for the uncorrected fluxes (\url{https://satdat.ngdc.noaa.gov/sem/goes/data/new_full/}); in the case of the Z89 intensities, the background counts are calculated through the comparison with uncorrected data. A 20\% systematical error is assumed for the Z89 fluxes (correction algorithm uncertainties). Statistical and systematical errors are summed in quadrature; only data with $<$100\% uncertainties are included.
\item The effective mean energy $E^{eff}_{i,mean}$ is determined by varying $E_{j}$ until $a_{i,j}$=1, which corresponds to the energy where, on average, EPEAD and PA\-ME\-LA intensities are equal: $F_{GOES,i}$ = $F_{PAM}\left(E^{eff}_{i,mean}\equiv E_{j}^{a_{i,j}=1}\right)$.
\item An effective range $\left(E^{eff}_{i,min}, E^{eff}_{i,max}\right)$ is associated to each channel by taking the energy values where $a_{i,j}$=1$\pm$0.5, i.e. $E^{eff}_{i,min}\equiv E_{j}^{a_{i,j}=0.5}$ and $E^{eff}_{i,max}\equiv E_{j}^{a_{i,j}=1.5}$. The effective ranges can be considered as
    an estimate of the derived mean energy uncertainties.
\end{itemize}

\begin{figure}[!t]
\centering
\includegraphics[width=5.2in]{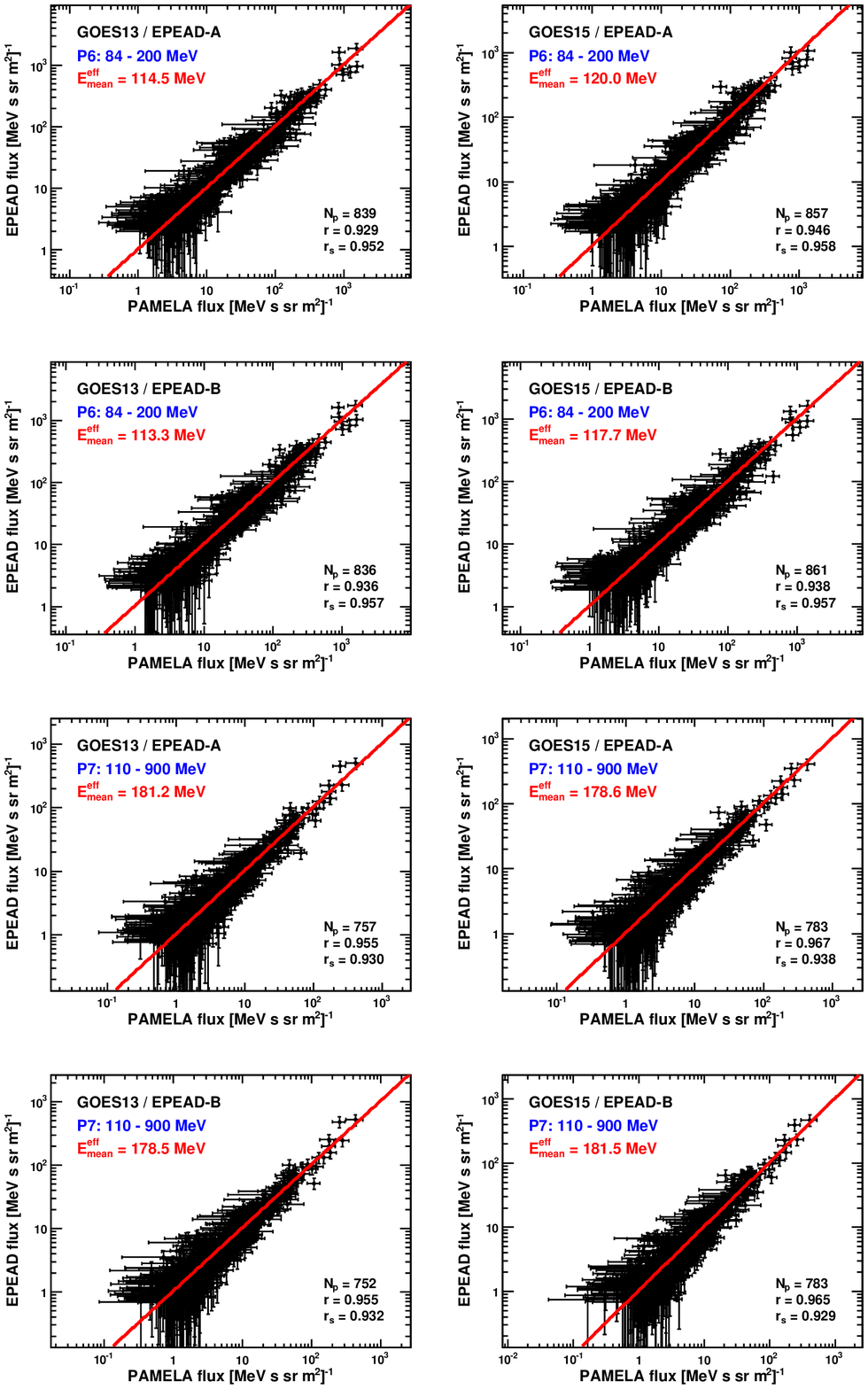}\\
\caption{\small{Uncorrected EPEAD versus PA\-ME\-LA differential flux distributions evaluated at the effective mean energies $E_{mean}^{eff}$ of the P6 and P7 channels, for GOES-13 (left panels) and GOES-15 (right panels). Results for the both the EPEAD-A and the EPEAD-B sensors are reported.
See the text for details.}}
\label{fig:G13G15epead_uncor}
\end{figure}

\begin{figure}[!t]
\centering
\includegraphics[width=5.2in]{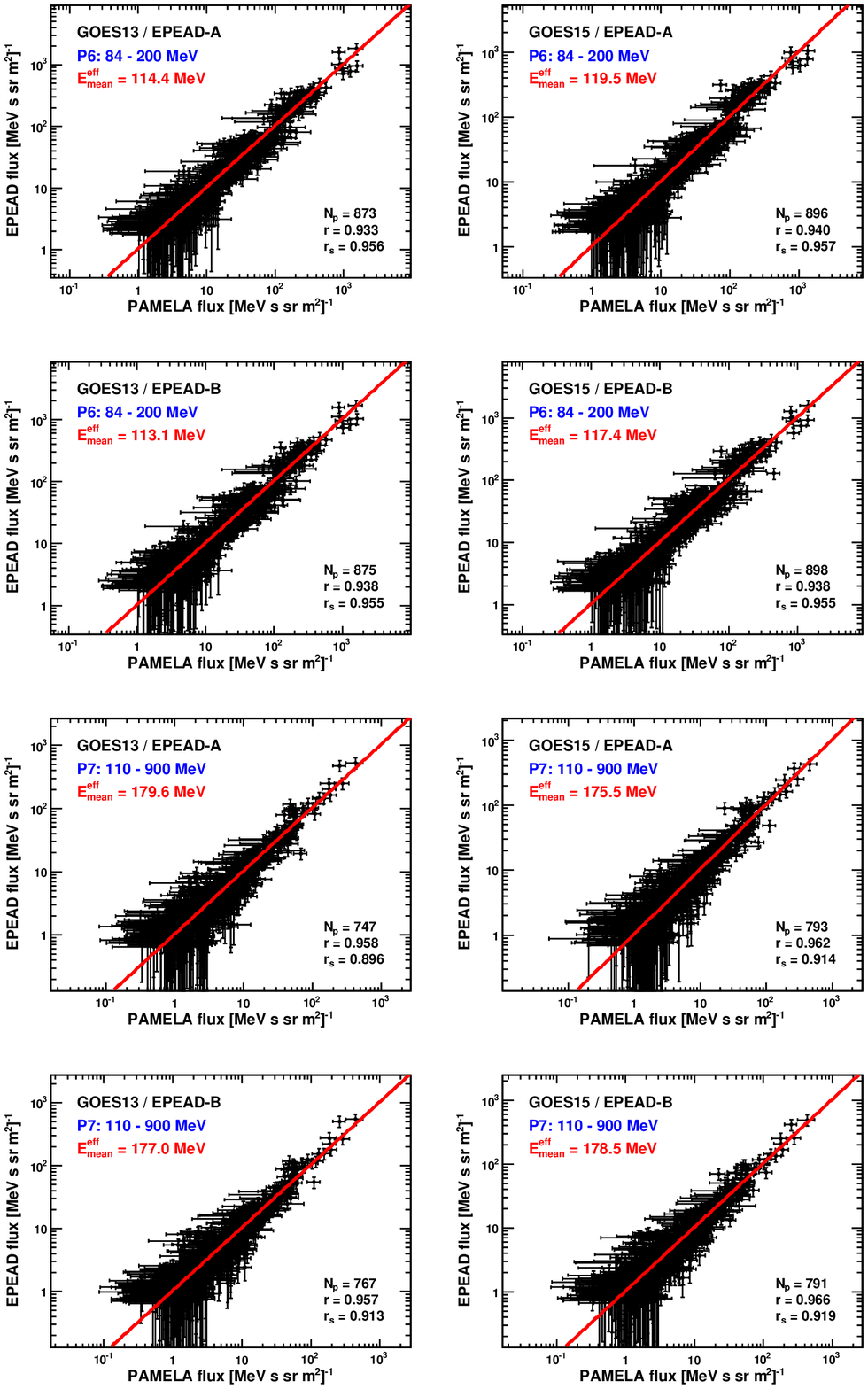}\\
\caption{\small{Same as Figure \ref{fig:G13G15epead_uncor}, but for Z89 EPEAD fluxes (with background subtracted).}}
\label{fig:G13G15epead_cor}
\end{figure}

The results of the calibration study are illustrated in Figures \ref{fig:G13G15epead_uncor} and \ref{fig:G13G15epead_cor}, where EPEAD versus PA\-ME\-LA differential fluxes at channel effective mean energies are shown for the uncorrected (with background subtracted) and the Z89 EPEAD intensities, respectively.
The P6 and P7 data are displayed for both EPEAD-A/B sensors;
left and right panels refer to GOES-13 and -15, respectively. The red lines correspond to $F_{GOES}=F_{PAM}$.
Each panel reports the nominal range, the derived effective mean energy $E^{eff}_{i,mean}$ and the number of data points used in the fit, along with the Pearson $r$ and the Spearman $r_{s}$ correlation coefficients. The horizontal and the vertical error bars accounts for the PA\-ME\-LA and the EPEAD flux uncertainties, respectively.

\begin{table}[!t]
\centering
\begin{tabular}{r||c|c||c|c||c|c||c|c}
\multicolumn{1}{r||}{} & \multicolumn{4}{c||}{GOES-13} & \multicolumn{4}{c}{GOES-15} \\
\hline
EPEAD sensor & \multicolumn{2}{c||}{A} & \multicolumn{2}{c||}{B} & \multicolumn{2}{c||}{A} & \multicolumn{2}{c}{B} \\
\hline
EPEAD channel & P6 & P7 & P6 & P7 & P6 & P7 & P6 & P7\\
\hline
E$^{eff}_{min}$(MeV) & 93.3 & 145.2 & 92.3 & 143.5 & 97.5 & 142.2 & 95.9 & 144.6 \\
E$^{eff}_{max}$(MeV) & 129.0 & 203.9 & 127.5 & 200.5 & 134.8 & 199.0 & 132.3 & 202.3 \\
E$^{eff}_{mean}$(MeV) & 114.4 & 179.6 & 113.1 & 177.0 & 119.5 & 175.5 & 117.4 & 178.5 \\
\hline
\end{tabular}
\caption{\small{Estimated effective ranges $\left(E^{eff}_{i,min}, E^{eff}_{i,max}\right)$ and mean energies $E_{mean}^{eff}$ of the EPEAD/P6-P7 channels, for both A/B sensors onboard GOES-13/15. Results are based on the uncorrected data.}}
\label{tab:epead_effective_energies_uncor}
\end{table}

\begin{table}[!h]
\centering
\begin{tabular}{r||c|c||c|c||c|c||c|c}
\multicolumn{1}{r||}{} & \multicolumn{4}{c||}{GOES-13} & \multicolumn{4}{c}{GOES-15} \\
\hline
EPEAD sensor & \multicolumn{2}{c||}{A} & \multicolumn{2}{c||}{B} & \multicolumn{2}{c||}{A} & \multicolumn{2}{c}{B} \\
\hline
EPEAD channel & P6 & P7 & P6 & P7 & P6 & P7 & P6 & P7\\
\hline
E$^{eff}_{min}$(MeV) & 93.3 & 146.7 & 92.4 & 144.6 & 97.9 & 145.0 & 96.1 & 147.3 \\
E$^{eff}_{max}$(MeV) & 129.1 & 205.6 & 127.8 & 202.5 & 135.3 & 202.3 & 132.8 & 205.7 \\
E$^{eff}_{mean}$(MeV) & 114.5 & 181.2 & 113.3 & 178.5 & 120.0 & 178.6 & 117.7 & 181.5 \\
\hline
\end{tabular}
\caption{\small{Same as Table \ref{tab:epead_effective_energies_uncor}, but based on the Z89 EPEAD data.}}
\label{tab:epead_effective_energies}
\end{table}

The channel effective mean energies along with associated effective ranges are summarized in Tables \ref{tab:epead_effective_energies_uncor} and \ref{tab:epead_effective_energies} for the uncorrected and the Z89 fluxes, respectively.
It can be noted that the two sets of results are in a good agreement for the P6 channel
while, for P7, the estimated Z89 energies are slightly higher;
these discrepancies can be attributed to differences regarding the Z89 background removal \citep{ref:Rodriguez2017}.
In general, the effective ranges are well contained in the nominal ranges (Table \ref{tab:nominal_ranges}).
Regarding the comparison between GOES-13 and -15 detectors, the energies of the latter are found to be slightly larger for all channels except for P7/A.

\subsection{HEPAD calibration}\label{HEPAD calibration}
The procedure employed to calibrate the four HEPAD energy channels (P8--P11) is analogous to the one developed for the EPEADs.
In this case, the calculation is significantly complicated by the limited statistics and by the background due to contaminations and GCRs.
Following the same approach used for the uncorrected EPEAD data,
the background is evaluated by using the average intensities registered by HEPAD
during the 24 hours prior to the SEP event onset.
The statistical errors are computed by using the 68.27\% confidence level intervals, taking into account the background subtraction.
Since the background is assumed constant during the SEP events, solar activity effects and other short timescale variations (including a diurnal modulation)
are ignored.
As discussed in Section \ref{PAMELA data}, the same background removal procedure is applied to the PAMELA data in order to minimize the uncertainties in the comparison.
Based on the observed discrepancies,
a 30\% systematical error is associated to the HEPAD points (mostly related to the background correction); only SEP data points with $<$100\% uncertainty are included.
Because of the small statistics, a 4-hour time resolution (corresponding to 5 PAMELA semi-orbits) is used for the comparison between PA\-ME\-LA and HEPAD fluxes.
The PA\-ME\-LA energy spectra are estimated by using 20 logarithmic bins covering the range 80 MeV -- 2 GeV and fitted by using the \citet{ref:ELLISON_RAMATY1995} function, accounting for the
expected ``roll-over'' in the high energy spectra \citep{ref:LEE2005} -- PAMELA SEP spectral observations will be presented and discussed in a future work.

The SEP events used for the calibration of each HEPAD channel can be inferred from the last column in Table \ref{tab:used_events}, which reports the highest GOES energy channel in which a discernible SEP signal was detected. It should be noted that the high-voltage settings on the GOES-13/15 HEPADs were increased on 2012 January 18 and 12, respectively, affecting the gain of the HEPAD photomultiplier tubes and, therefore, the channel energies as well as the accuracy of measured particle \citep{ref:Rodriguez2012}. Consequently, the data points of the five SEP events prior to 2012 January (see Table \ref{tab:used_events}) are excluded from the calculation.

\begin{figure}[!t]
\centering
\includegraphics[width=5.2in]{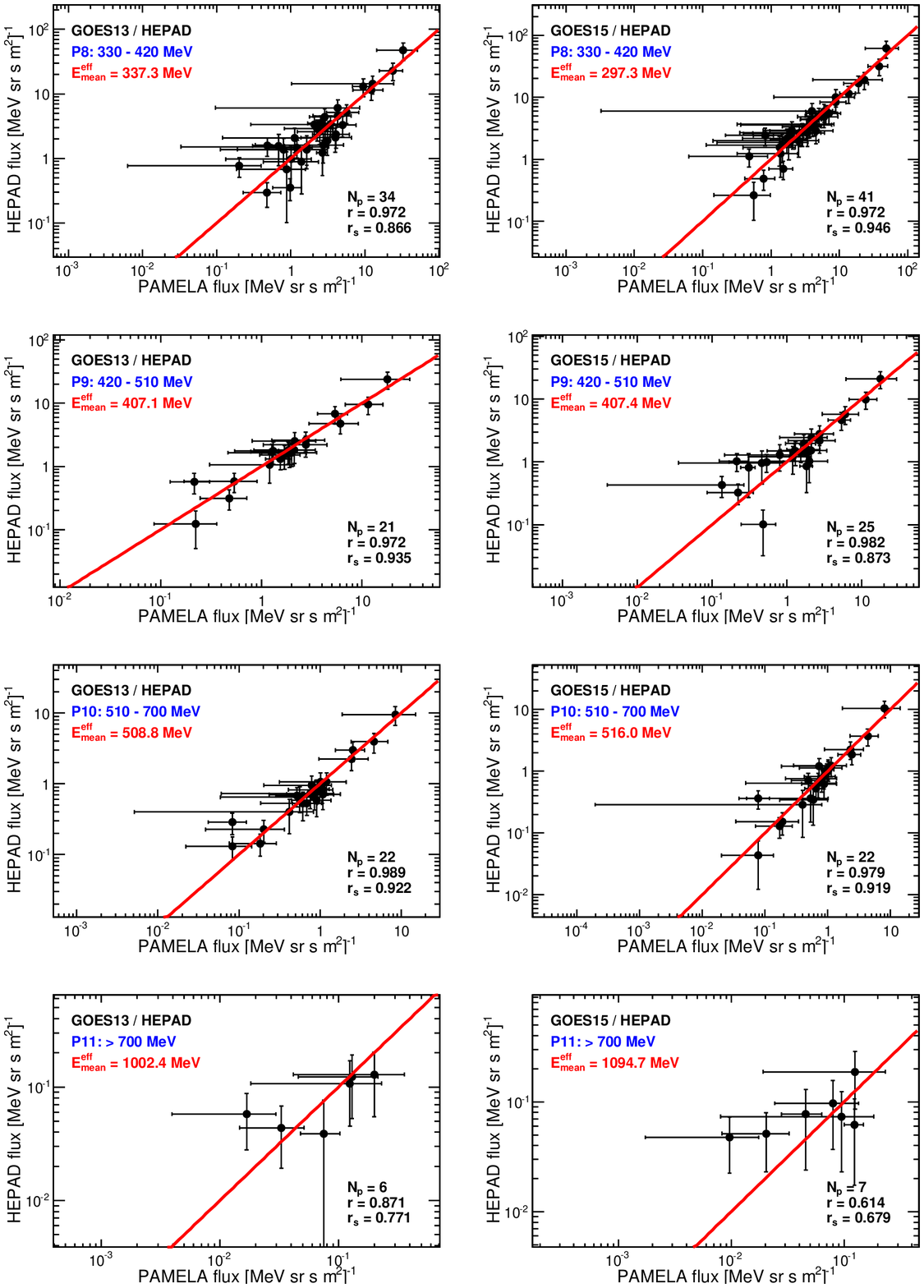}\\
\caption{\small{Background-corrected HEPAD versus PA\-ME\-LA differential flux distributions evaluated at the effective mean energies $E_{mean}^{eff}$ of the four HEPAD channels (P8--P11), for GOES-13 (left panels) and GOES-15 (right panels). See the text for details.}}
\label{fig:HepadGOES}
\end{figure}

The results of the HEPAD calibration study are displayed in Figures \ref{fig:HepadGOES} for both GOES-13 (left panels) and -15 (right panels). For each channel (P8--P11), the PA\-ME\-LA versus HEPAD differential flux distributions at derived mean energies are shown; the red lines correspond to $F_{GOES}=F_{PAM}$. Each panel reports the corresponding nominal range, the evaluated mean energy $E^{eff}_{i,mean}$ and the number of data points used in the fit, along with the Pearson $r$ and the Spearman $r_{s}$ correlation coefficients. The horizontal and the vertical error bars accounts for the PA\-ME\-LA and the HEPAD flux uncertainties, respectively.

\begin{table}[!t]
\centering
\begin{tabular}{r||c|c|c|c||c|c|c|c}
\multicolumn{1}{r||}{} & \multicolumn{4}{c||}{GOES-13} & \multicolumn{4}{c}{GOES-15} \\
\hline
HEPAD channel & P8 & P9 & P10 & P11 & P8 & P9 & P10 & P11 \\
\hline
E$^{eff}_{min}$(MeV) & 273.9 & 330.0 & 418.7 & 852.6 & 240.4 & 325.3 & 420.4 & 878.6 \\
E$^{eff}_{max}$(MeV) & 387.5 & 458.0 & 566.0 & 1081.2 & 335.6 & 464.6 & 573.1 & 1230.0 \\
E$^{eff}_{mean}$(MeV) & 337.3 & 407.1 & 508.8 & 1002.4 & 297.3 & 407.4 & 516.0 & 1094.7 \\
\hline
\end{tabular}
\caption{\small{Effective ranges $\left(E^{eff}_{i,min}, E^{eff}_{i,max}\right)$ and mean energies $E_{mean}^{eff}$ computed for the four HEPAD channels (P8--P11, background-corrected intensities) onboard GOES-13 and -15.}}
\label{tab:hepad_effective_energies}
\end{table}

Table \ref{tab:hepad_effective_energies} summarizes the channel effective mean energies along with the associated effective ranges, computed
as described in the previous Section.
The effective energy of the P8 channel is found to be higher for GOES-13, while the P10--P11 energies are higher for GOES-15; for P9 the estimated values are similar.
It should be noted that, in contrast to the EPEADs, the effective ranges of the HEPAD channels span energies lower than the nominal intervals and, in particular, the effective energies of GOES-13/P9-P10 and of GOES-15/P8-P9 are outside the nominal ranges, suggesting possible absolute calibration discrepancies.
On the other hand, it is remarkable and reassuring that the calculated P11 mean energies are consistent with the 1000 MeV value assumed in the channel data processing.

\section{Discussion}\label{Discussion}
\begin{figure}[!t]
\centering
\includegraphics[width=0.85\textwidth]{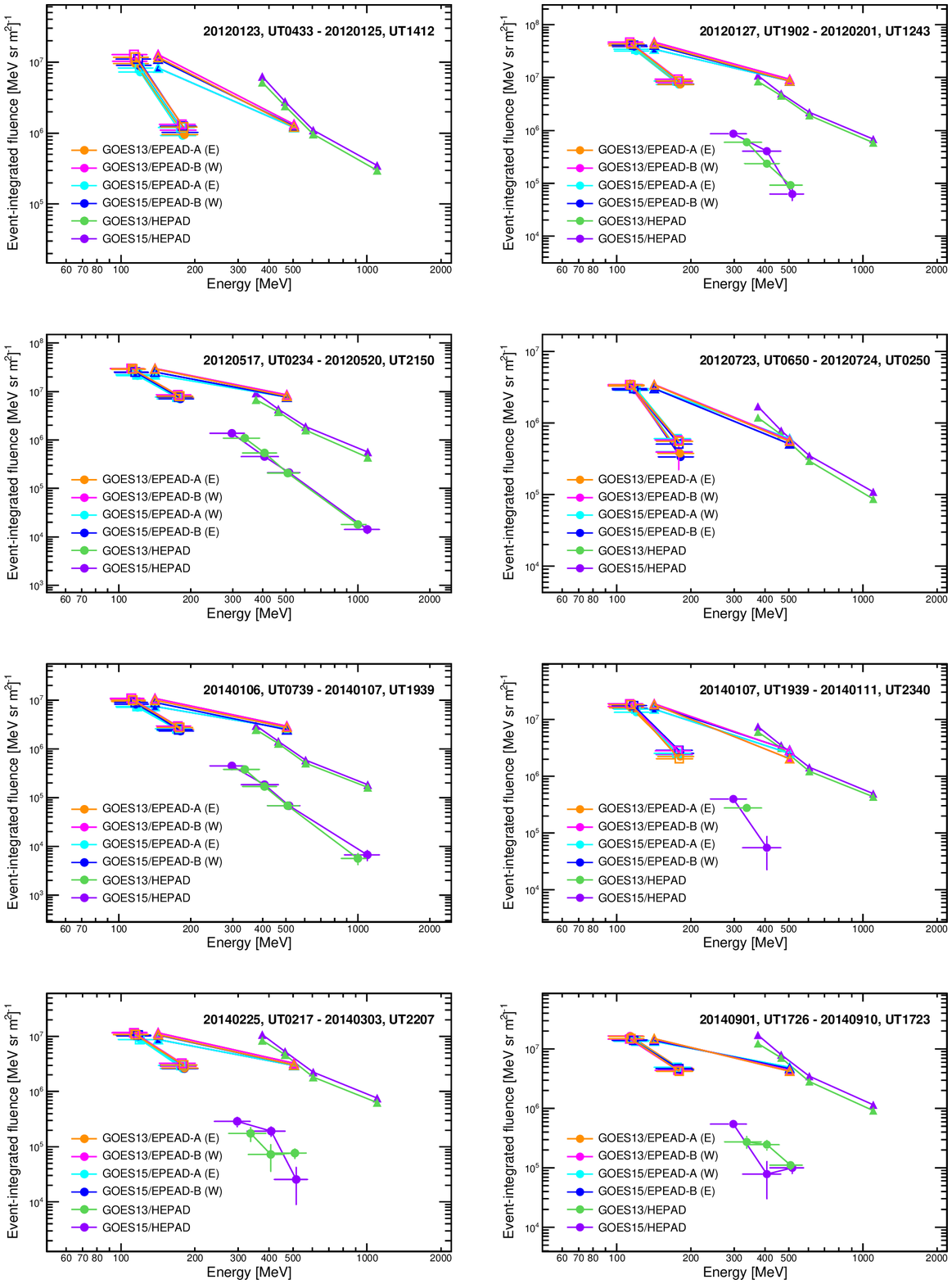}
\caption{\small{Event-integrated proton fluences measured by GOES-13/15 above 80 MeV, during 8 SEP events (see the dates reported in each panel). The triangles represent the original EPEAD and HEPAD points (channels: P6--P11) evaluated by using the nominal mean energies (see Table \ref{tab:nominal_ranges}) and including the background. The full circles refer to the calibrated EPEAD (based on uncorrected data) and HEPAD points, with the background subtracted; for a comparison, calibrated EPEAD points obtained with the Z89 data are also reported (empty squares). See the text for details.}}
\label{fig:fluences_comp}
\end{figure}

Figure \ref{fig:fluences_comp} displays the event-integrated fluences $\int F(t) dt$ measured by the GOES-13/15 proton detectors, during 8 selected SEP events.
The start/stop dates, based on PAMELA data (48-min resolution) and fixed for all energies, are reported in each panel. The triangles represent the original EPEAD and HEPAD points (channels: P6--P11) evaluated by using the nominal mean energies (see Table \ref{tab:nominal_ranges}) and including the background. The full circles refer to the calibrated EPEAD (based on uncorrected intensities) and HEPAD points, with the background subtracted; for a comparison, calibrated EPEAD points obtained with the Z89 data are also shown (empty squares). Lines are to guide the eye. The vertical bars include the statistical uncertainties (including the background subtraction), while the horizontal bars represent the derived effective ranges. The orientation (E/W) of the EPEAD-A/B sensors is also indicated in the legends.

A good agreement between the two EPEAD calculations can be observed for most events (points are superimposed), while the Z89 intensities (especially for P7) appear to be relatively higher during small events (e.g. 2012 January 23 and 2012 July 23); however, reported differences are
compatible with estimated errors.
The improvement with respect to the nominal GOES data is evident both for the EPEAD and the HEPAD channels. In particular, measurement uncertainties are significantly reduced for P7, and the HEPAD results are consistent with an extrapolation of EPEAD points at higher energies. The large correction in HEPAD data is mostly due to the background removal; fluences are null for the two aforementioned events. Regarding the P11 channel, only the two registered GLEs contribute to an observable SEP signal.

\begin{figure}[!t]
\centering
\includegraphics[width=0.9\textwidth]{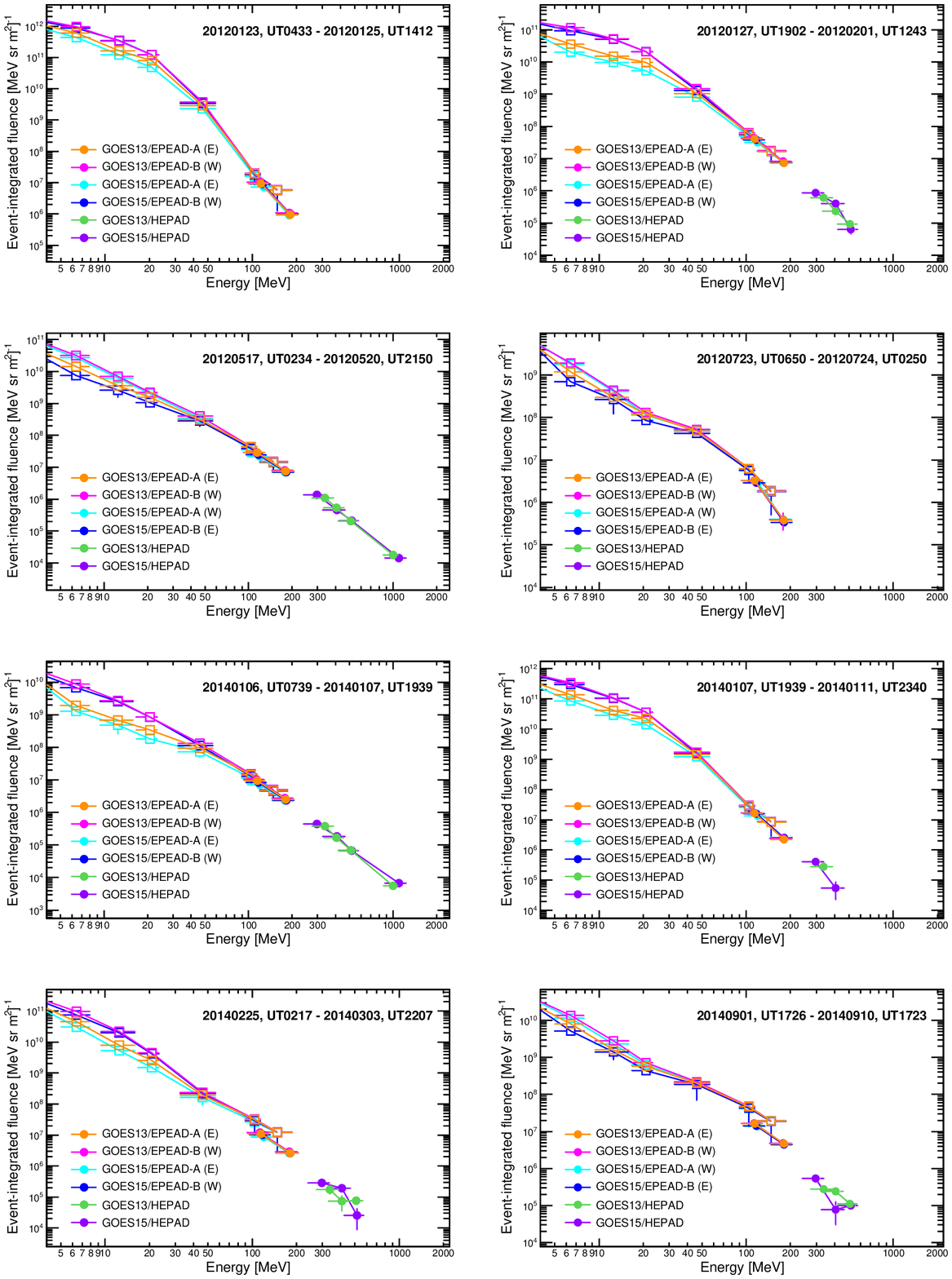}
\caption{\small{Event-integrated fluences measured by GOES-13/15 above 5 MeV, during the same 8 events as Figure \ref{fig:fluences_comp}. The background-removed GOES points derived in this work ($E$$\gtrsim$100 MeV, full circles)
are compared to the EPEAD data based on the results by \citet{ref:Sandberg2014} ($E$$\lesssim$200 MeV, empty squares).
See the text for details.}}
\label{fig:fluences}
\end{figure}

Finally, Figure \ref{fig:fluences} shows the comparison with the uncorrected (including background) EPEAD fluences based on the S14 calibrated energies (channels: P2--P7, derived for GOES-11/EPS), denoted by empty squares; in this case, associated vertical error bars refer to the differences with respect to the Z89 intensities. Lines are to guide the eye.
The P6 and P7 points are in a good agreement for most intense SEP events, while some discrepancies can be noted for relatively small events. This is not unexpected since the background is included in S14,
consequently the corresponding mean energies are lower ($E_{mean,6}^{eff}$=104 MeV and $E_{mean,7}^{eff}$=148 MeV), and the resulting fluences appear to be overestimated (especially for the P7 channel) when the background is not negligible.
On the other hand, the two calculations appear to be consistent
when the background correction is accounted for, as can be inferred from the comparison with the Z89 data (S14 point error bars).
Concerning the low energy channels below PA\-ME\-LA threshold (P2--P5),
the differences between the flux intensities measured by east/westward looking detectors can be ascribed to East-West effects \citep{ref:Rodriguez2010}. As discussed in Section \ref{Section:GOES data}, such effects are suppressed at PA\-ME\-LA energies.

\section{Summary and conclusions}
Thanks to the superior energy resolution and the wide explored interval, PAMELA observations offer an unique opportunity to calibrate the high energy (E$>$80 MeV) proton detectors onboard GOES-13 and -15.
The effective energies of the P6--P11 channels, determined through a systematical comparison with PAMELA SEP measurements, are presented in this work.
In particular, the response of the two highest energy channels (P6--P7) of the EPEADs (including both A and B sensors) is investigated for both uncorrected and Z89 data sets. Mostly important, the calibration study is extended to the four HEPAD channels (P8--P11)
operating at higher energies.

Suggested corrections significantly reduce the uncertainties on the proton intensities provided by GOES instruments, especially the HEPADs, improving the reliability of the differential energy spectra measured during SEP events. More sophisticated calibration approaches
will be implemented
and discussed in forthcoming publications.

\section*{Acknowledgments}
The author gratefully thanks the PA\-ME\-LA collaboration for the use of the proton data,
and acknowledges very helpful discussions with J.~V. Rodriguez and I. Sandberg about the GOES-EPEAD/HEPAD observations and related calibration studies.
He would also like to thanks M. Martucci for the support with the PA\-ME\-LA measurements,
and E.~R. Christian, G.~A. de Nolfo, I.~G. Richardson and J.~M. Ryan for constructive comments.

\end{document}